\documentclass[aps,prl,twocolumn,showpacs,superscriptaddress]{revtex4-1} 
\usepackage{epsfig,amsmath,amssymb,bm,epsf,graphicx,psfrag}
\usepackage[all]{xy}
\usepackage{graphicx}
\usepackage{color}
 
\pdfoutput=1

\newcommand{\ket}[1]{|{#1}\rangle}

\newcommand{\clus}{|{C}\rangle}

\newtheorem{Theorem}{Theorem}

\newtheorem{Lemma}{Lemma}

\newtheorem{Prop}{Proposition}

\begin{document}

\title{A computationally universal phase of quantum matter}

\author{Robert Raussendorf}
\affiliation{Department of Physics and Astronomy, University of British Columbia, Vancouver, BC, Canada}
\affiliation{Stewart Blusson Quantum Matter Institute, University of British Columbia, Vancouver, BC, Canada}

\author{Cihan Okay}
\affiliation{Department of Physics and Astronomy, University of British Columbia, Vancouver, BC, Canada}
\affiliation{Stewart Blusson Quantum Matter Institute, University of British Columbia, Vancouver, BC, Canada}

\author{Dong-Sheng Wang}
\affiliation{Department of Physics and Astronomy, University of British Columbia, Vancouver, BC, Canada}
\affiliation{Stewart Blusson Quantum Matter Institute, University of British Columbia, Vancouver, BC, Canada}

\author{David T. Stephen}
\thanks{These authors contributed equally to this work}
\affiliation{Max-Planck-Institut f{\"u}r Quantenoptik, Hans-Kopfermann-Stra{\ss}e 1, 85748 Garching, Germany}

\author{Hendrik Poulsen Nautrup}
\thanks{These authors contributed equally to this work}
\affiliation{Institut f{\"u}r Theoretische Physik, Universit{\"a}t Innsbruck, Technikerstr. 21a, A-6020 Innsbruck, Austria}

\date{\today}

\begin{abstract}
We provide the first example of a symmetry protected quantum phase that has universal computational power. This two-dimensional phase is protected by one-dimensional line-like symmetries that can be understood in terms 
of local symmetries of a tensor network. These local symmetries imply that every ground state in the phase is a
universal resource for measurement based quantum computation.
\end{abstract}


\maketitle

In the presence of symmetry, quantum phases of matter can have computational power. This was first conjectured in \cite{Doh}-\cite{Bartl}, and has been proven \cite{MM2}-\cite{SPTO1} or numerically supported \cite{Darmawan},\cite{HuaWei} in several instances. The important property is that the computational power is uniform. It does not depend on the precise choice of the state within the phase, and is thus a property of the phase itself. In this way, phases of quantum matter acquire a computational characterization and computational value. 

The quantum computational power of physical phases is utilized by measurement based quantum computation (MBQC) \cite{RB01}, where the process of  computation is driven by local measurements on an initial entangled state. Here, we consider initial states that originate from symmetry protected topological (SPT) phases \cite{GW}-\cite{Wen2}. 

Proofs of the existence of such  ``computational phases of quantum matter'' have so far been confined to spatial dimension one. After it was shown that computational wire---the ability to shuttle quantum information from one end of a spin chain to the other---is a property of certain SPT phases \cite{Bartl}, the first phase permitting quantum computations on a single logical qubit was described in \cite{MM2}. In fact, uniform computational power is ubiquitous in one-dimensional SPT phases \cite{SPTO2}, \cite{SPTO1}.

Computationally, physical phases in dimension 2 and higher are more interesting than in dimension 1. The reason is that, in MBQC, one spatial dimension plays the role of circuit model time. Therefore, MBQC in dimension $D$ corresponds to the circuit model in dimension $D-1$, and universal MBQC is possible only in $D\geq 2$.

Yet, to date, the evidence for quantum computational phases of matter is much more scant for $D\geq 2$ than for $D=1$. Numerical evidence exists for deformed Affleck-Kennedy-Lieb-Tasaki Hamiltonians on the honeycomb lattice \cite{Darmawan, Zitz, HuaWei}. In addition, extended regions of constant computational power have also been observed in SPT phases with  $\mathbb{Z}_2$-symmetry \cite{WeiHua}.

Numerous computationally universal resources states for MBQC have been constructed \cite{Poul}--\cite{MM3} using the tools of group cohomology that also form the basis for the classification of SPT order \cite{Wen1}, \cite{Wen2}. From the starting point of these special states, it remains open what happens to the computational power as one probes deeper into the SPT phases surrounding them. 

For the cluster phase in $D=2$, a symmetry protected phase that contains the cluster state, it was shown analytically that universal computational power persists throughout 
a finite region around the cluster state \cite{Bartl2}.

\begin{figure}[hb!]
\begin{center}
\includegraphics[width=6.5cm]{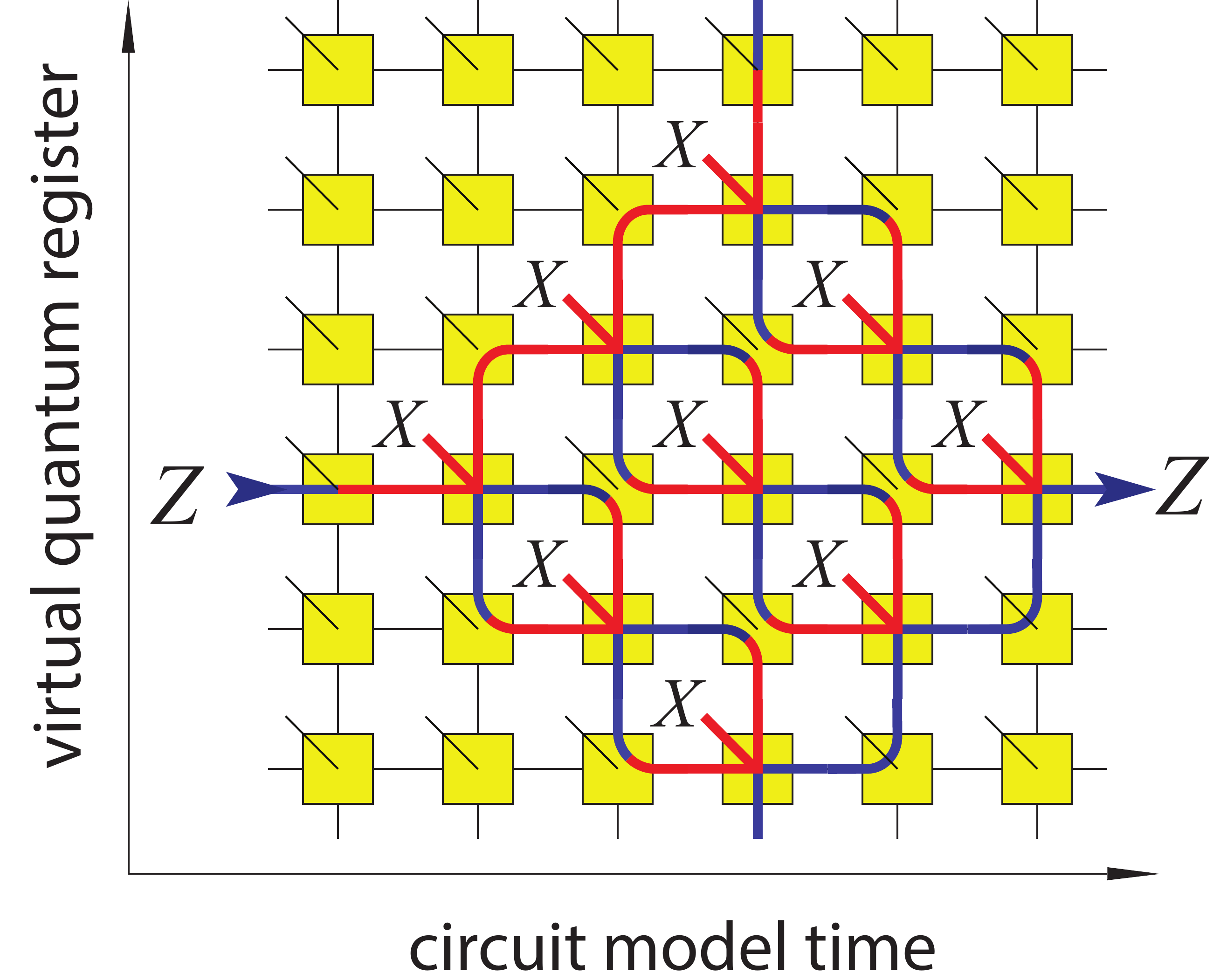}
\caption{\label{QCAf} Symmetry protected quantum correlations enable uniform computational power throughout the 2D cluster phase. The long-range symmetry shown is composed of the symmetries of local PEPS tensors.}
\end{center}
\end{figure}

Here, we prove the existence of a computationally universal phase of quantum matter in spatial dimension two. As in \cite{Bartl2}, the phase we consider is protected by one-dimensional line-like symmetries, generalizing the conventional notion of symmetry protected topological order defined by global on-site symmetries. 
As in the case of global symmetries, these line symmetries can be built from the local symmetries of a tensor network which
persist throughout the phase. Using this, we establish that computational universality persists throughout the entire phase. The backbone of the computational scheme is symmetry protected correlations in a virtual quantum register, see Fig.~\ref{QCAf}. \medskip

\begin{figure}
\begin{center}
\begin{tabular}{lcl}
(a)\\
\multicolumn{3}{l}{\includegraphics[width=6.8cm]{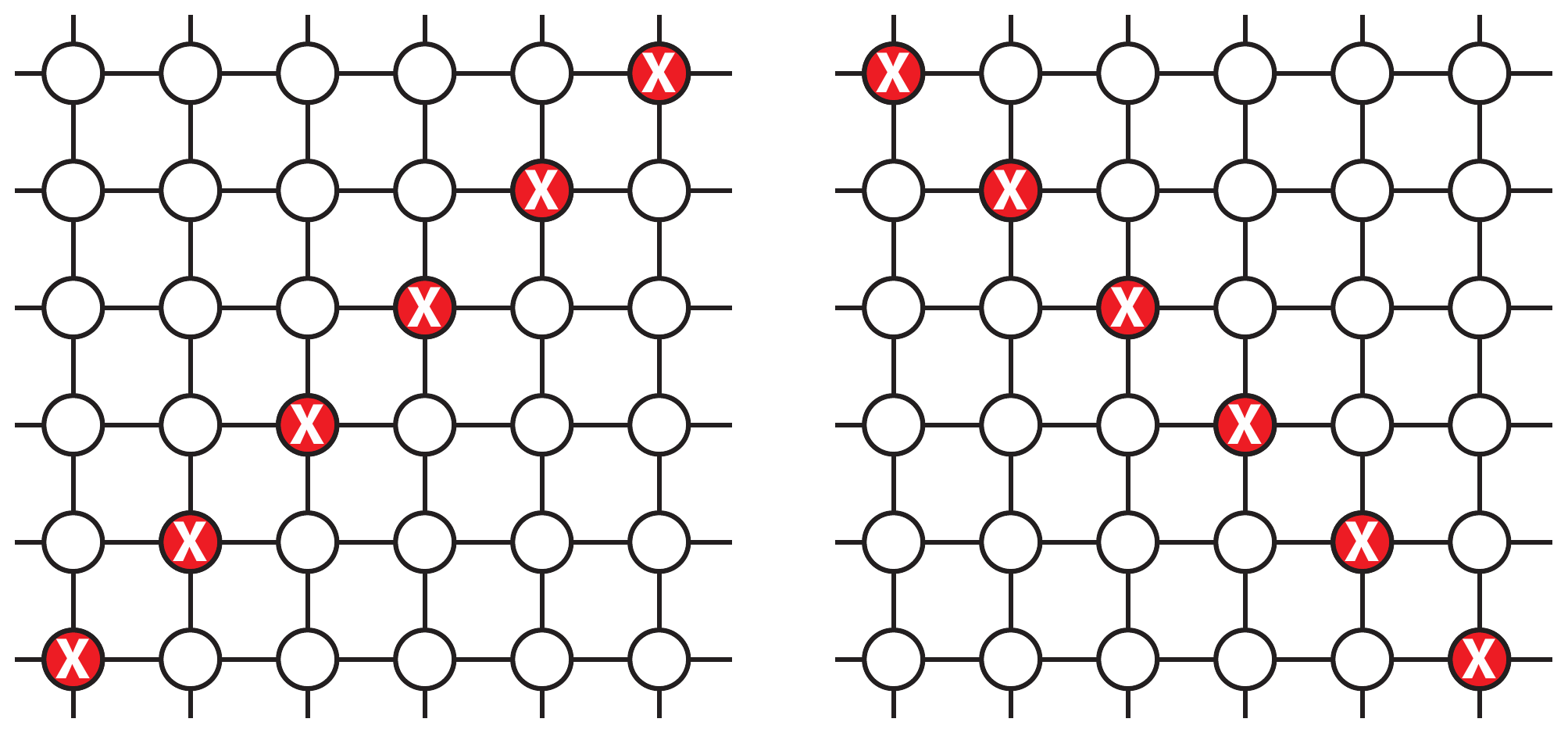}}\\
(b) && (c)\\
\includegraphics[width=3.0cm]{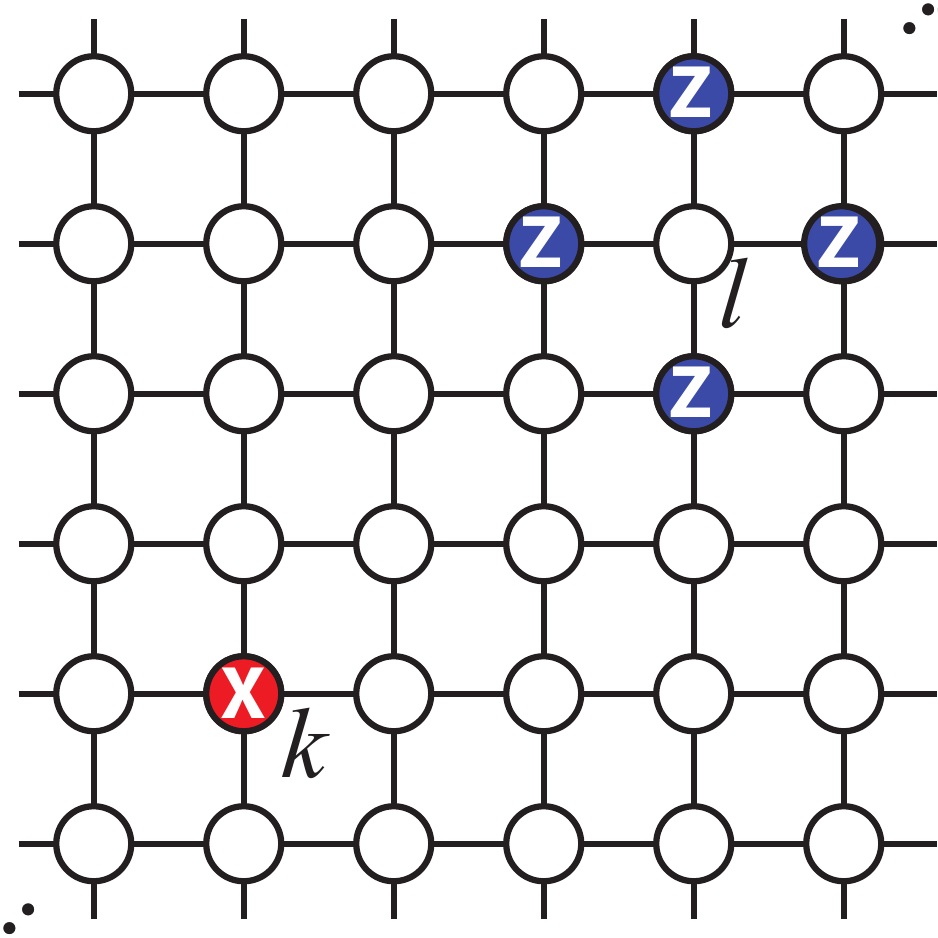} && \includegraphics[width=4.4cm]{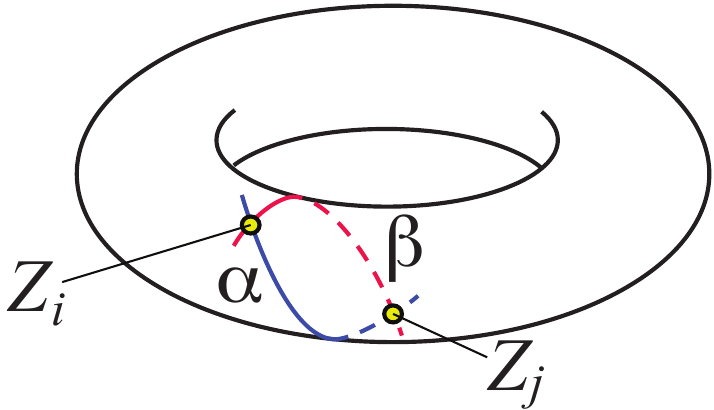}
\end{tabular}
\caption{\label{Stripes} (a) Stripe-like symmetry of Eq.~(\ref{StripeSymm}). All translates are also symmetries. (b)-(c) The generators of Pauli operators that commute with the symmetries Eq.~(\ref{StripeSymm}).  (b) The local operators $X_k$ and $\text{Star}_l$, for all sites $k,l$.  (c) Geometrically non-local operators $Z_i\otimes Z_j$. The locations $i$, $j$ are consecutive intersections of the supports $\alpha$, $\beta$ of two symmetries.}
\end{center}
\end{figure}

{\em{Setting and result.}} We consider a two-dimensional (2D) simple square spin lattice which is, for simplicity of boundary conditions, embedded in a torus of a small circumference $n$ and a large circumference $nN$, with $n,N\in \mathbb{N}$, $N\gg n$ and $n$ even. Its Hamiltonian is invariant under all lattice translations and the symmetries
\begin{equation}\label{StripeSymm}
U_{c,+} =\bigotimes_{x=0}^{nN-1} X_{x,c+x},\; U_{c,-} =\bigotimes_{x=0}^{nN-1} X_{x,c-x},
\end{equation}
for all $c \in \mathbb{Z}_n$. Therein, $X\equiv\sigma_x$, and the addition in the second index of $X$ is mod $n$.  A graphical rendering of these symmetries is provided in Fig.~\ref{Stripes}a. These symmetries were previously considered in \cite{Bartl2}. We consider phases in which the ground state is unique, and thus shares the symmetries. 

As the Hamiltonian is varied while respecting the symmetries Eq.~(\ref{StripeSymm}), the respective ground states arrange into phases. The central object of interest is the ``2D cluster phase'', i.e. the physical phase which respects the symmetries Eq.~(\ref{StripeSymm}) and which contains the 2D cluster state.

The main result of this paper is the following.
\begin{Theorem}\label{Tm}
For a spin-1/2 lattice on a torus with circumferences $n$ and $Nn$, where $N\gg n$ and $n$ even, all ground states in the cluster phase, except a possible set of measure zero,  are universal resources for measurement-based quantum computation on $n/2$ logical qubits.
\end{Theorem}

To facilitate the proof of Theorem~\ref{Tm}, we introduce the notion of ``cluster-like'' states $|\Phi\rangle$. For  a square grid in dimension 2, we represent states as projected entangled pair states (PEPS) with local tensors $A_\Phi$, such that contracting virtual legs on a torus as in Fig.~\ref{QCAf} describes the wave function of $\ket{\Phi}$. The ``cluster-like'' states are those whose PEPS tensors have the symmetries
\begin{equation}\label{2Dsymm}
\parbox{7.8cm}{\includegraphics[width=7.7cm]{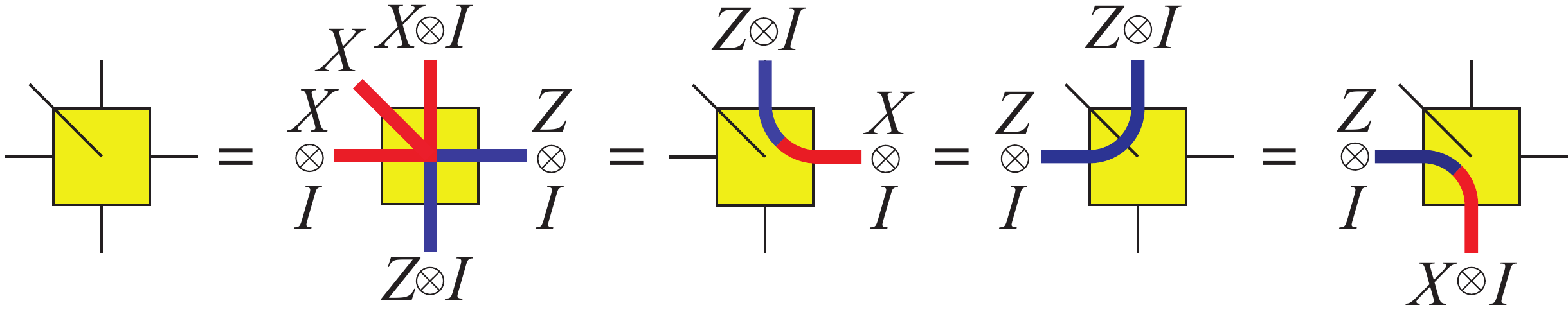}}
\end{equation}
Therein, red (blue) legs indicate Pauli operators $X$ ($Z$). This notation means that, for example, acting on the physical leg (diagonally directed) of the PEPS tensor with a Pauli $X$ is equivalent to a corresponding action of Pauli operators on the virtual legs. The reason for calling states satisfying Eq.~(\ref{2Dsymm}) ``cluster-like'' is that if we add 
$$
\parbox{3cm}{\includegraphics[width=3cm]{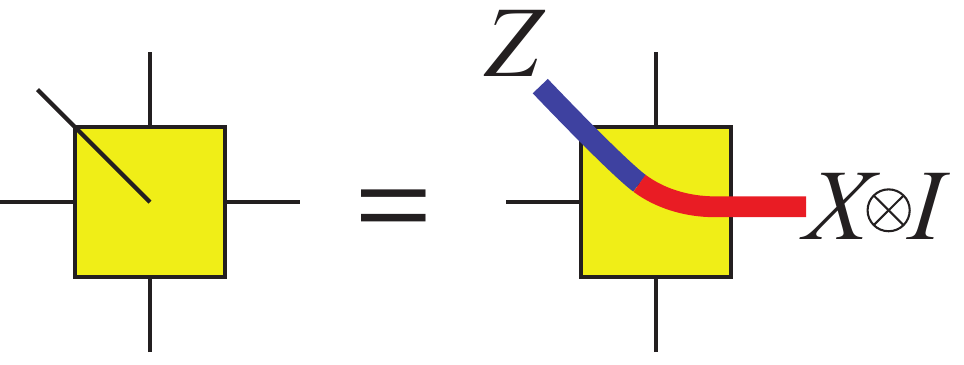}}
$$
to those symmetries, then we obtain cluster states as the only solution of the joint symmetry constraints. 

The proof of Theorem~\ref{Tm} splits into two parts. First we show that all states in the 2D cluster phase are cluster-like, and then demonstrate that cluster-likeness implies universal computational power.
\medskip

{\em{A 2D physical phase of cluster-like states.}} Here we prove the following result.
\begin{Prop}\label{2Dphase}
Every ground state $|\Phi\rangle$ in the 2D cluster phase has a description in terms of a local tensor $A_\Phi$ that has the symmetries of Eq.~(\ref{2Dsymm}). 
\end{Prop}
Our starting point is the characterization of SPT phases in terms of symmetric quantum circuits. A symmetric quantum circuit is a sequence of unitary gates $U = \prod_{i=1}^l U_i$ where each gate $U_i$ is invariant under the symmetry group $G$ of Eq.~(\ref{StripeSymm}),  $[U_i,U(g)] = 0$, for all $g \in G$. In a local such circuit, each gate $U_i$ acts only on a bounded number of qubits \cite{HuangChen}. We then have the following result \cite{Wen1}, 
\begin{Lemma}\label{SRC} 
Symmetric gapped ground states in the same SPT phase are connected by symmetric local quantum circuits of constant depth.
\end{Lemma}
To prove Proposition~\ref{2Dphase}, we analyze the structure of the symmetry-respecting gates $U_{\Phi,i}$ of the circuit $U_\Phi$ mapping the cluster state $|C\rangle$ to a given state $|\Phi\rangle$ in the cluster phase, $|\Phi\rangle = U_\Phi |C\rangle$. Writing
\begin{equation}\label{Uphi}
U_{\Phi,i} = \sum_j d_j P_j,\;\text{with}\; d_j \in \mathbb{C},\;\forall j,
\end{equation}
only symmetry-respecting $n$-qubit Pauli operators $P_j$, $P_j\in {\cal{P}}_n$, can appear on the r.h.s. The generators of such Pauli operators are displayed in Fig.~\ref{Stripes} (b), (c). Furthermore, the operators shown in Fig.~\ref{Stripes} (c) do not contribute since they are geometrically non-local. Thus, the Pauli operators appearing on the r.h.s. of Eq.~(\ref{Uphi}) are generated by local operators $X_k$ and $Z$-type star operators $\text{Star}_l$, for all sites $k,l$ of the lattice (See Section I A of the Supplementary Material (SM), Lemma~3).

Now expanding the entire circuit $U_\Phi$ into a sum of Pauli operators, every Pauli operator in this expansion is also a product of $X_k$ and Star-operators. We further observe that, by the form of the cluster state stabilizer,
\begin{equation}\label{Star1}
\text{Star}_k \clus = X_k\clus,
\end{equation}
for all lattice sites $k$. Using relation Eq.~(\ref{Star1}), all star operators in the expansion of $U_\Phi$ can be eliminated. We thereby obtain a transformation $T_\Phi$ that satisfies the relation $T_\Phi\clus = U_\Phi\clus = |\Phi\rangle$, and is composed of Pauli-$X$ operators only,
\begin{equation}\label{Exp2}
T_\Phi=\sum_\textbf{j} c_\textbf{j} X(\textbf{j}).
\end{equation}
Therein, $X(\textbf{j}):=\bigotimes_k (X_k)^{j_k}$ an $X$-type Pauli operator with support on the $n\times nN$ torus, i.e., $\textbf{j}$ is a binary vector with $n^2N$ components. 
\medskip

{\em{Proof of Proposition~\ref{2Dphase}.}} To illustrate the idea of the proof, we first discuss the special case where the map $T_\Phi$ is a tensor product of local factors, $T_\Phi=\bigotimes_{k} t_{\Phi,k}$. Then, to obtain a local tensor $A_\Phi$ representing $|\Phi\rangle$, we apply $T_\Phi$ site-wise to the local tensor $C$ representing the cluster state. Graphically,
$$
\parbox{2.9cm}{\includegraphics[width=2.8cm]{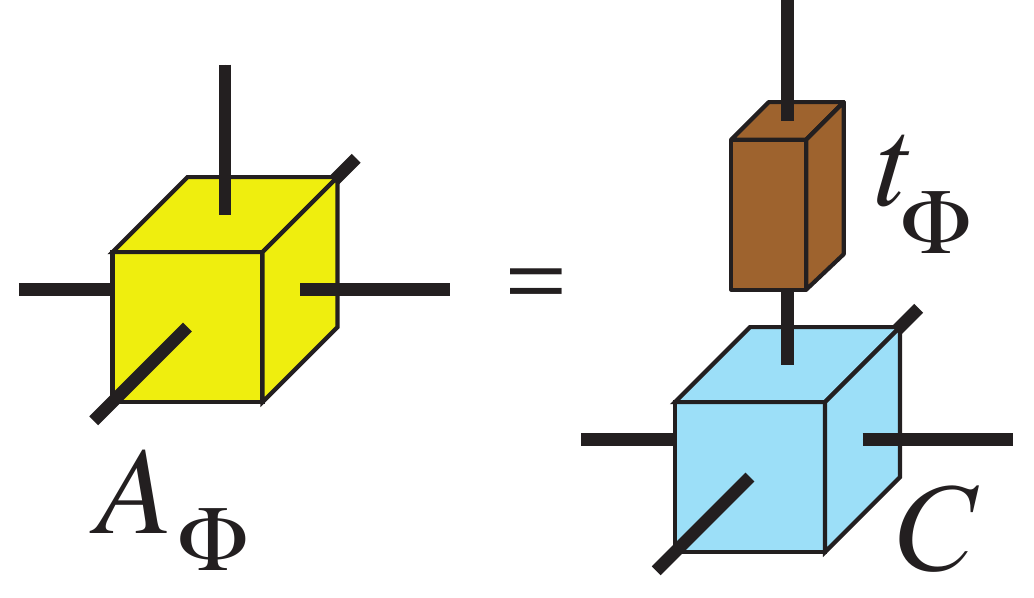}}.
$$
Since by Eq.~(\ref{Exp2}) $t_\Phi$ is a linear combination of $I$ and $X$, it commutes with $X$.
Hence, the symmetries Eq.~(\ref{2Dsymm}) of the cluster state tensors $C$ are also symmetries of the tensors $A_\Phi$ representing $|\Phi\rangle$. \medskip

Now turning to the general case, the action of $T_\Phi$ on $|C\rangle$ results in local tensors $A_\Phi$ of the form
\begin{equation}\label{LTb}
\parbox{3.0cm}{\includegraphics[width=2.8cm]{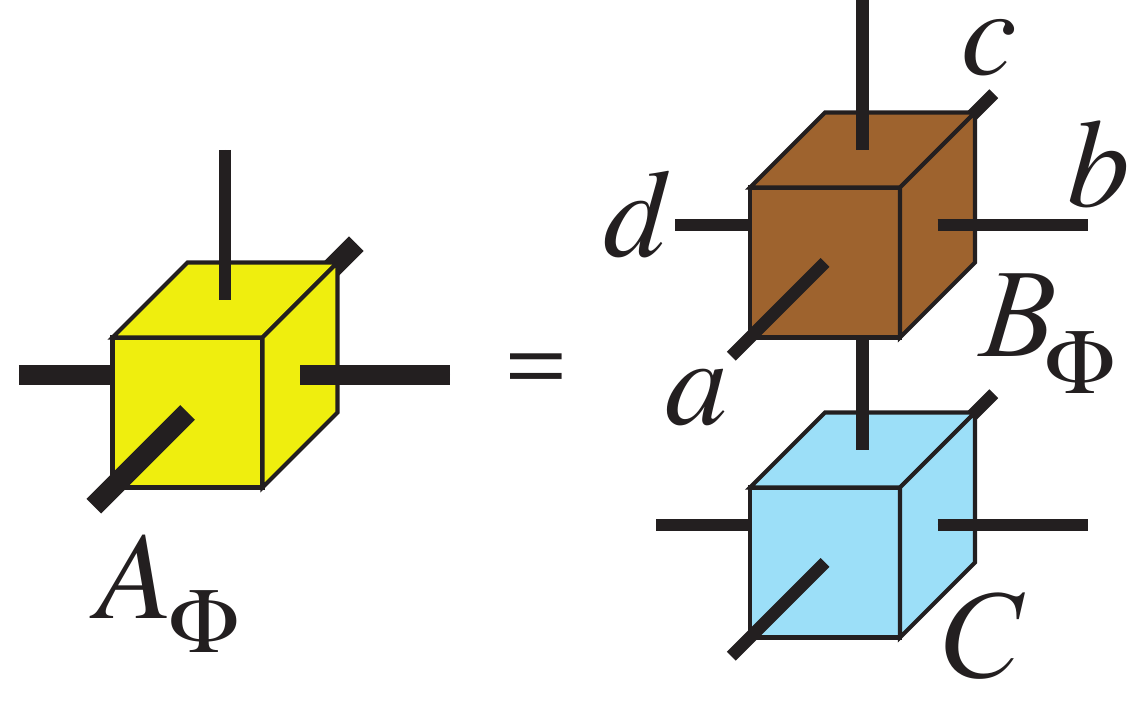}},
\end{equation}
where the ``junk tensor'' $B_\Phi$ \cite{Bartl} forms a tensor network representation of the map $T_\Phi$, and emerges as a consequence of the non-locality of the map $T_\Phi$. It inherits from $T_\Phi$ the property that on the physical leg of $C$ (pointing upwards) it acts as $I$ or $X$, depending on the state of the virtual links $a,..,d$ (for details, see the SM, Section I B). The junk tensor $B_\Phi$ thus commutes with the action of the local Pauli $X$-operator, 
$$
\parbox{3cm}{\includegraphics[width=3cm]{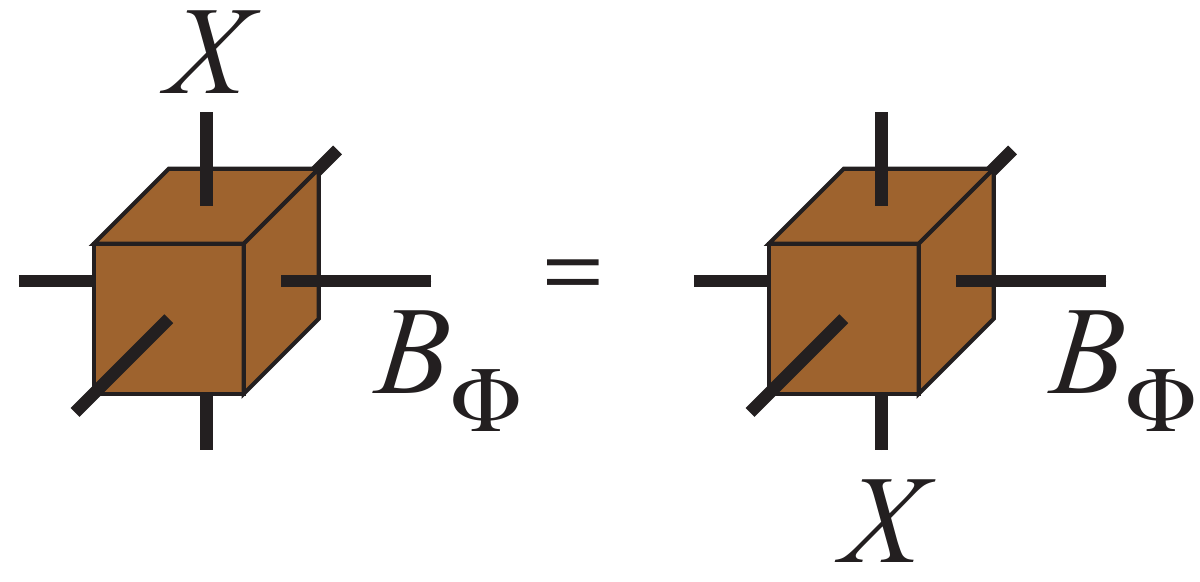}}.
$$
In result, the symmetries Eq.~(\ref{2Dsymm}) hold for all tensors $A_\Phi$ describing a state $|\Phi\rangle$ in the cluster phase. $\Box$
\medskip

{\em{Cluster symmetries and computation.}} We now show that the symmetries Eq.~(\ref{2Dsymm}) of PEPS tensors imply MBQC universality of the corresponding quantum state. 
This proceeds in two steps. We establish (i) computational wire, i.e. the ability to shuttle quantum information across the torus, and (ii) a universal set of quantum gates.

(i) {\em{Computational wire.}} We now map to a quasi-1D setting by grouping spins into blocks of size $n\times n$. If we block $n\times n$ copies of the tensor $A_\Phi$, as in Fig. \ref{QCAf}, we obtain the block tensor $\mathcal{A}_\Phi$ which forms a matrix product state (MPS) representation of the quasi-1D system. Contracting the physical legs of this tensor with local $X$ eigenstates labelled by the $n^2$-component binary vector $\mathfrak{i}$ gives the tensor component $\mathcal{A}_\Phi(\mathfrak{i})$. We can now use the symmetries in Eq.~(\ref{2Dsymm}) to constrain these tensor components:

\begin{Lemma}\label{QCA}
	Consider a torus of size $n\times nN$, and $n\in 2\mathbb{N}$. 
	For all ground states $|\Phi\rangle$ in the 2D cluster phase, the corresponding block tensors $\mathcal{A}_\Phi(\mathfrak{i})$ satisfy 
	\begin{equation}\label{BlockSplit}
	\mathcal{A}_\Phi(\mathfrak{i})=\mathcal{C}(\mathfrak{i})\otimes \mathcal{B}_\Phi(\mathfrak{i}).
	\end{equation} 
	The logical tensors $\mathcal{C}(\mathfrak{i})$ are constant throughout the phase, and
	\begin{equation}\label{refocus}
	\mathcal{C}(\mathfrak{i}) \in {\cal{P}}_n,\;\;\forall \mathfrak{i}.
	\end{equation}
\end{Lemma}
Lemma~\ref{QCA} establishes the primitive of computational wire, similar to Theorem 1 in [3]. The Hilbert space on which the tensor components $\mathcal{A}_\Phi(\mathfrak{i})$ act is the so-called virtual space, which decomposes into ``logical subsystem" and  "junk subsystem" [3]. Upon measurement in the $X$-basis of all spins in a block, the logical subsystem is acted on by the operators $\mathcal{C}(\mathfrak{i})$, which are uniform across the cluster phase. Conversely, the operators $\mathcal{B}_\Phi(\mathfrak{i})$ acting on the junk space vary uncontrollably across the phase. Thus, to achieve computation, the logical subspace is used to encode and process information. The operators $\mathcal{C}(\mathfrak{i})$ become the usual outcome-dependent byproduct operators of MBQC. They are of computational use, as described below under ``quantum gates".

Two points are worth noting, one technical, one physical. (i) With Lemma~\ref{QCA}, we have mapped the original two-dimensional system to an effectively one-dimensional system composed of blocks. A wealth of techniques established for 1D SPT order thereby becomes available \cite{Bartl}--\cite{SPTO1}, \cite{GW}--\cite{Wen2}. (ii) The blocking notwithstanding, the basis $\{|\mathfrak{i}\rangle\}$ in which Eq.~(\ref{refocus}) holds is local {\em{at the level of individual spins}}, not only at block level. (It is the local $X$-eigenbasis.) Since MBQC uses 1-spin local measurements, we require this stronger notion of locality.

Finally, we explain why Lemma~\ref{QCA} is a consequence of the symmetries of the local tensors $A_\Phi$ in the cluster phase.  The local symmetries Eq.~(\ref{2Dsymm}) can be combined in such a way that they map Pauli operators on the virtual logical register one column farther to the right,
\begin{equation}\label{ring}
	\parbox{6.4cm}{\includegraphics[width=6.4cm]{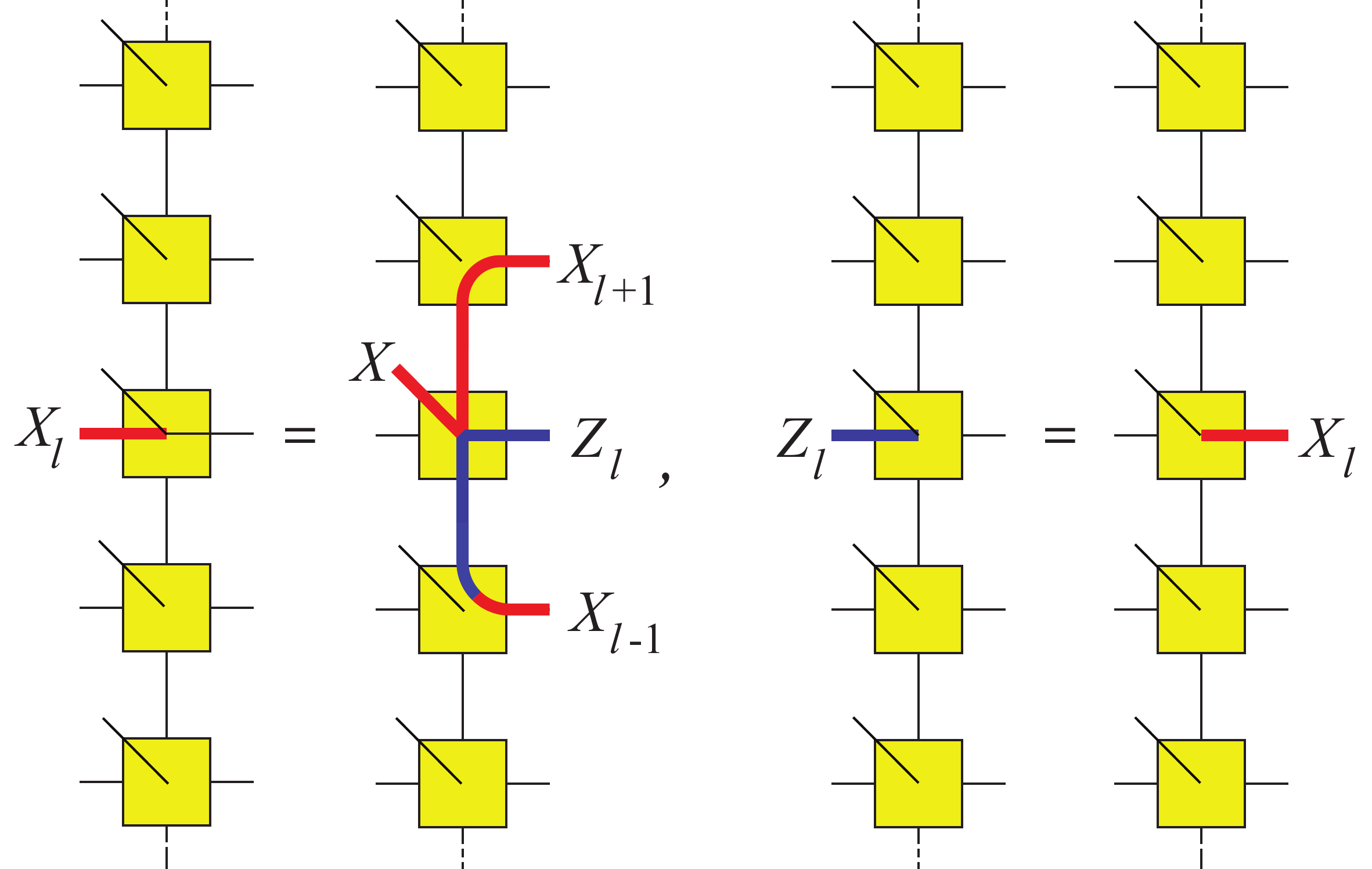}},
\end{equation}
for all $l$. (The tensor factors ``$I$'' for the action of the symmetries on the junk systems have been omitted). Iterating these propagation relations $n$ times ($n$ is the circumference of the torus), we find that, upon measurement of the physical qubits in the local $X$-basis, each virtual local Pauli operator $Z$ is mapped onto itself up to sign. See Fig.~\ref{QCAf} for illustration ($n=6$ is shown). The same is true for Pauli operators $X$, cf. Fig.~4 in the SM. Thus, every virtual Pauli operator is mapped to itself up to sign, after one clock cycle of duration $n$. Therefore, the action of $\mathcal{A}_\Phi$ on the logical subsystem is indeed by Pauli operators, as stated by Lemma~\ref{QCA}. As a technical remark, we note that the following construction requires that Lemma~\ref{QCA} holds also when $\mathcal{A}_\Phi$ is put into the so-called canonical MPS form \cite{MPS}. Details of this condition, as well as the proof of its veracity, are given in the SM, Section III A.\smallskip

(ii) {\em{Quantum gates.}} The subsequent construction significantly differs from the standard mapping to the circuit model \cite{RB01}. Specifically, the technique of ``cutting out coupled wires'' by local $Z$-measurements is not available throughout the cluster phase, and is therefore replaced.

As a first step, we observe that the byproduct operators $\mathcal{C}(\mathfrak{i})$ are of the form
\begin{equation}\label{Cprod}
\mathcal{C}(\mathfrak{i}) \sim \prod_{k \in K} \mathcal{C}[k]^{i_k},
\end{equation}
where ``$\sim$'' is equality up to phase, $K$ is the $n\times n$ block of spins, and $i_k$ the measurement outcome at location $k$.

Eq.~(\ref{Cprod}) means that every site $k$ in the block has its own byproduct operator $\mathcal{C}[k]$. This is known to hold for the cluster state \cite{RB01}, and by Lemma~\ref{QCA} it extends to the entire cluster phase.

Next, we find the precise form of the byproduct operators $\mathcal{C}[k]$ for certain sites $k\in K$. Namely, for the sites $k=(1,l)$, $(2,l)$ and $(n,l)$ in the first, second and last column of each block, the operators $\mathcal{C}[k]$ are
\begin{equation}\label{QCAsum}
\begin{array}{rcl}
\mathcal{C}[(1,l)]&=&Z_l,\\
\mathcal{C}[(2,l)]&=&Z_{l-1}X_lZ_{l+1},\\
\mathcal{C}[(n,l)]&=&X_l.
\end{array}
\end{equation}
They can be understood as follows. For the last column in the block, $n$, the operator $\mathcal{C}[n,l]$ is the standard byproduct operator for cluster states. By Lemma~\ref{QCA} it holds in the entire cluster phase. (See the SM, Sec. III B for the result in canonical form.) The ${\cal{C}}[r,l]$ for earlier columns $r$ are also $X$-operators, inserted at position $(r,l)$. They are then propagated forward to the right boundary of the block using Eq.~(\ref{ring}), resulting in Eq.~(\ref{QCAsum}). 

If the resource is a 2D cluster state, the special state in the phase of interest, then onsite measurements in the $X/Y$-plane of the Bloch sphere are universal \cite{Fitz}. Because of the product form of the byproduct operators Eq.~(\ref{Cprod}), every local measurement implements one logical gate. Suppose the measurement at site $k$ is in the basis spanned by
$|0,\alpha\rangle_k = \cos(\alpha) \, |0\rangle_k - i\sin(\alpha)\, |1\rangle_k$,  
$|1,\alpha\rangle_k = -i\sin (\alpha) \, |0\rangle_k + \cos(\alpha)\, |1\rangle_k$,
with $|0\rangle$, $|1\rangle$ referring to eigenstates of $X$.
The resulting gate is $U_\alpha(i_k) = \mbox{}_k\langle i_k,\alpha|0\rangle_k \,I + \mbox{}_k\langle i_k,\alpha|1\rangle_k \, \mathcal{C}[k]$, hence
$$
U_\alpha(i_k) = \mathcal{C}[k]^{i_k} \exp(i\alpha \, \mathcal{C}[k]).
$$
Here, the operators $\mathcal{C}[k]$ of Eq.~(\ref{refocus}) become a computational tool, as it specifies the unitary gate implemented.
The outcome-dependent byproduct operator can be compensated for by classical side-processing and adaptive measurement bases \cite{RB01}. With Eq.~(\ref{QCAsum}), the gate set 
\begin{equation}
\mathcal{U} = \{e^{i\alpha\, Z_{l-1}X_lZ_{l+1}}, e^{i\alpha \, Z_l}, e^{i\alpha \, X_l},\; \forall \alpha \in \mathbb{R}\}.
\end{equation}
can be realized. $\mathcal{U}$ is  a universal set \cite{TIQC}; also see Section IV B of the SM.

When moving away from the cluster state into the cluster
phase, non-trivial tensors $\mathcal{B}_\Phi$ appear, and measurement in a local basis away from the symmetry-respecting $X$-basis becomes non-trivial. If unaccounted for, the logical subsystem becomes entangled with the junk subsystem through such measurement \cite{Bartl}, which introduces decoherence into the logical processing. However, this undesirable effect can be prevented by the techniques of \cite{SPTO1}.
By virtue of Lemma~\ref{QCA}, we mapped to a quasi-1D setting to which we can apply Theorem~2 of \cite{SPTO1}. (The essentials of \cite{SPTO1} are reviewed in Section~IV A of the SM.) In result, the universal gate set $\mathcal{U}$ can be implemented in the whole cluster phase, not only on the cluster state.
\smallskip

To summarize, the argument for computational universality of the 2D cluster phase splits into two parts. First, we have shown in Proposition~\ref{2Dphase} that all ground states in the 2D cluster phase are cluster-like, i.e., they satisfy the symmetry constraints Eq.~(\ref{2Dsymm}).  Second, by mapping to a quasi one-dimensional system we showed that the symmetries Eq.~(\ref{2Dsymm}) lead to universal computational power. Taken together, these two results yield Theorem~\ref{Tm}.

\medskip 

{\em{Conclusion.}} We have described the first symmetry protected topological phase in which every ground state (up to a possible set of measure zero) has universal power for measurement based quantum computation. Our phase is protected by symmetries acting on a lower dimensional subsystem, and it is associated with a set of local symmetries of tensor networks, see Eq.~(\ref{2Dsymm}). These symmetries are sufficient to guarantee computational universality of the corresponding tensor network. What implications these symmetries have on the physics of this phase and others like it remains an interesting question. 

As for the implications on the computational side, we ask: Can the computational power of quantum phases of matter be classified? In the spirit of this question, we conclude with three more specific ones: (i) How broadly can the present construction be generalized? (ii) The line-like symmetries we consider are neither global symmetries, which are typically used to define SPT phases, nor are they local like in a lattice gauge theory. Indeed, they are more closely related to the ``higher-form" symmetries considered in \cite{Yosh},  \cite{Bartl3}, \cite{KT} which act on lower dimensional submanifolds of the whole lattice. Is this type of symmetry necessary for a computationally universal phase, or can other structurally different symmetries lead to similar results? (iii) As one-dimensional computational phases \cite{SPTO2},\cite{SPTO1} build on symmetry protected computational wire \cite{Bartl}, the present construction builds on a symmetry protected quantum cellular automaton. In particular, Eq.~(\ref{ring}) defines the transition function of a quantum cellular automaton. Quantum cellular automata have been classified \cite{W1}-\cite{W4}. What is the relation between this classification and computational phases of quantum matter?\smallskip

\begin{acknowledgments}
This work is supported by NSERC (CO, DSW, DTS, RR), Cifar (RR), the Stewart Blusson Quantum Matter Institute (CO and DSW), and the Austrian Science Fund FWF within the DK-ALM: W1259-N27 (HPN). DSW thanks Z.C. Gu for discussions. 
\end{acknowledgments}

\newpage

\begin{widetext}
\begin{center}
\textbf{SUPPLEMENTARY MATERIAL}\vspace{2mm}\\

\textbf{for}\vspace{2mm}\\

\textbf{\large{A computationally universal phase of quantum matter}}
\end{center}
\end{widetext}

\refstepcounter{figure}\refstepcounter{figure}
\refstepcounter{Lemma}\refstepcounter{Lemma}

\noindent
{\em{Remark regarding notation:}} Numbered equations from the main text are labeled with a suffix ``[m]''. For example, Eq.~(3) from the main text is referenced in this Supplementary Material as ``Eq.~(3) [m]''. The numbering of figures and lemmas is continued from the article.

\section{Proof of Proposition 1} 

To complete the proof of Proposition~1 [m] we need to show (a) that all bounded range Pauli operators commuting with the symmetries Eq.~(1) [m] are products of local $X$-operators and $Z$-type star operators, and (b) that any state $|\Phi \rangle$ in the cluster phase can be expressed in terms of local tensors $A_\Phi$ of the form Eq.~(6) [m]. The respective arguments are provided in Sections~\ref{pepSy} and \ref{GenA} below. 
 
\subsection{Symmetric short-range circuits}\label{pepSy}

\begin{Lemma}\label{LocProd}
For a torus of size $n\times nN$, any tensor product of Pauli $Z$-operators that is symmetric under the transformations Eq.~(1) [m] and whose support fits into a skewed square of horizontal and vertical extension smaller than $n$ is equal to a product of star operators.
\end{Lemma}

To prove Lemma~\ref{LocProd} we need a further result.
\begin{Lemma}\label{Reloc}
Any product of $Z$-operators within a $k\times k$ region of a skewed square lattice can be moved to the boundary of that region by multiplying with star operators.
\end{Lemma}

\begin{figure}
\begin{center}
\begin{tabular}{ll}
(a) & (b) \\
\includegraphics[height=3.9cm]{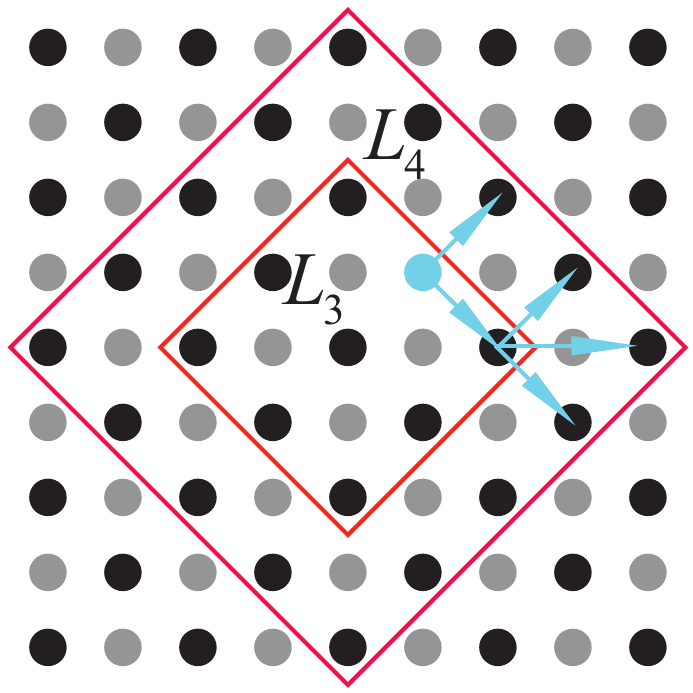} & \includegraphics[height=3.9cm]{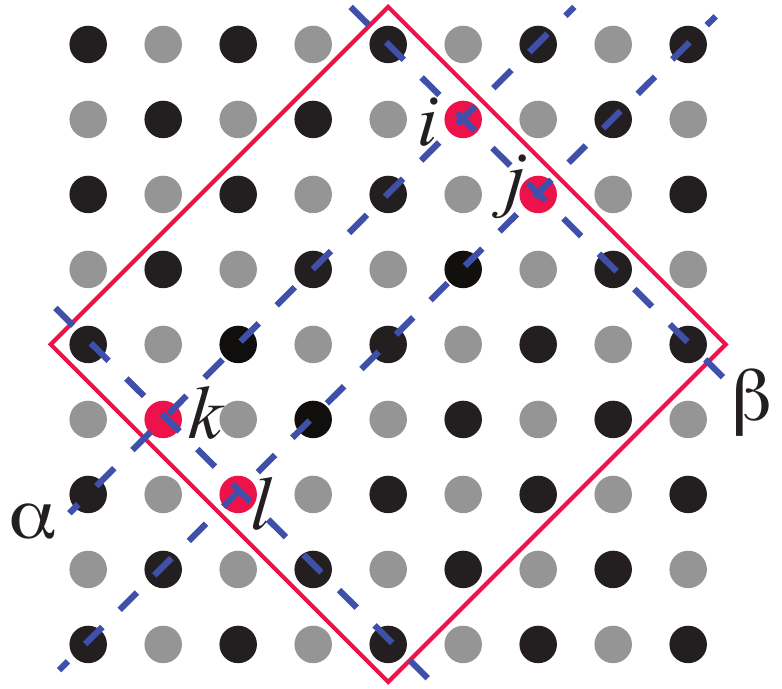}
\end{tabular}
\caption{\label{Z_Illu} (a) Illustration for Lemma~\ref{Reloc}. Relocating Pauli operators $Z$ to the boundary of the skewed square by multiplying with star operators. (b) Illustration for Lemma~\ref{LocProd}. By the symmetries Eq.~(1) [m], every Pauli operator $Z_i$ in the boundary of a small skewed square region has 3 three distinct partners.}
\end{center}
\end{figure}

{\em{Proof of Lemma~\ref{Reloc}.}} W.l.o.g., consider one of the two sub-lattices (even or odd). The proof is by induction. (I) The statement is true for a skewed square $L_2$ of size $2\times 2$. We now prove that (II) If the statement is true for the skewed square $L_{k-1}$, then it is also true for $L_k$. W.l.o.g., assume that $Z$-operators are only located on the boundary of $L_{k-1}$. Each such $Z$ can be moved into a corner of $L_k$ by multiplying with star operators, leaving a trail of $Z$s in the boundary of $L_k$; see Fig.~\ref{Z_Illu}a.  $\Box$\medskip 

{\em{Proof of Lemma~\ref{LocProd}.}} W.l.o.g., consider one of the two sub-lattices. By Lemma~\ref{Reloc}, the $Z$-operators can all be moved to the boundary of that region by multiplying with star operators. Since the initial operator $Z(\textbf{v})$ is invariant under the transformations Eq.~(1) [m], and the star operators are invariant, so is the resulting operator $Z(\textbf{w})$. Now consider a tensor factor $Z_i$ in $Z(\textbf{w})$. Since $Z(\textbf{w})$ is symmetric under the transformations (1) [m] and these symmetries act on diagonals, it exists in conjunction with three additional local operators $Z_j$, $Z_k$, $Z_l$; see Fig.~\ref{Z_Illu}b for the labeling. The two diagonals $\alpha$, $\beta$ that intersect in $i$ only intersect again on the torus at a horizontal and vertical distance of $n/2$ (see Fig.~2b). Since by assumption the support of $Z(\textbf{w})$ is contained within a skewed square of vertical and horizontal extension smaller than $n$, $\alpha$ and $\beta$ do not intersect twice within such a square. Therefore, the locations $i$, $j$, $k$, $l$ are all distinct, and 
their product can  be removed by a product of star operators. The procedure is iterated until no local Pauli operators $Z$ remain. $\Box$

\subsection{Structure and symmetry of the tensors $A_\Phi$}\label{GenA}

Here we prove Eq.~(6) [m], with the additional symmetry property $B_\Phi X = X B_\Phi$. We already sketched the argument in the main text, and now provide additional detail.

We denote the 5-legged PEPS tensors of a cluster state (one physical, four virtual legs) by $C$; also see \cite{Eis}. The corresponding tensors for the other states $|\Phi\rangle=U_\Phi \clus$ in the cluster phase are denoted by $A_\Phi$. We prove that the tensor $A_\Phi$ is invariant under the symmetries Eq.~(2) [m], by constructing it explicitly. 

With each transformation $T_\Phi$ we associate a network of ``junk'' tensors that has the same geometry as the tensor network for $\clus$. That is, at each lattice site resides a tensor $B_\Phi$ with four virtual legs and two physical legs (input and output). The Hilbert space associated to each virtual leg is spanned by the vectors $|\textbf{j},a,b\rangle$, where $a\in[1,nN]$ and $b\in[1,n]$ are integers and $\textbf{j}$ is the same index appearing in Eq.~(5) [m], which we rewrite here for convenience:
\begin{equation} \label{tphi}
 T_\Phi=\sum_{\textbf{j}} c_\textbf{j} X(\textbf{j}).
\end{equation}
Each pair $a,b$ naturally corresponds to a site in the 2D lattice, which we denote $k_{a,b}$.  

We define the tensor $B_\Phi$ by the following non-zero components,
\begin{equation} \label{Localtens4}
\parbox{3.5cm}{\includegraphics[width=3.5cm]{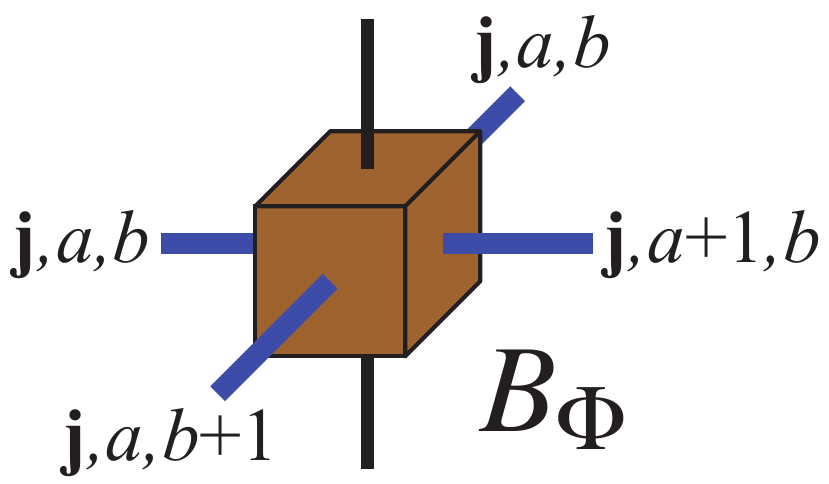}} = c_\textbf{j}^{\frac{1}{n^2N}}\parbox{1cm}{\includegraphics[width=1cm]{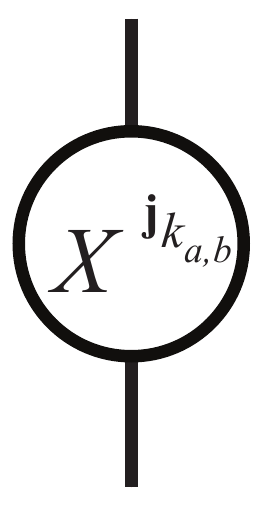}}.
\end{equation}
We now need to establish two properties of the tensor $B_\Phi$, namely (i)
The transformation $T_\Phi$ corresponds to the tensor network composed of $B_\Phi$, with all virtual legs pairwise contracted, and (ii) The tensor $A_\Phi$ representing a state $|\Phi\rangle$ satisfies the symmetries of Eq.~(2) [m].

(i) First note that $T_\Phi$ is invariant under all lattice translations: for every $\textbf{j}$, all operators that can be obtained from $X(\textbf{j})$ via lattice translations appear with the same coefficient $c_\textbf{j}$ in Eq.~(\ref{tphi}). Then we can rewrite,
\begin{equation} \label{tphi2}
 T_\Phi=\frac{1}{n^2N} \sum_{\textbf{j}} c_\textbf{j} [X(\textbf{j})+\mathrm{trans.}],
\end{equation}
where trans. indicates all possible lattice translations of $X(\textbf{j})$, which may or may not be distinct from $X(\textbf{j})$. In order to obtain Eq.~(\ref{tphi2}) from Eq.~(\ref{Localtens4}) We contract the network of tensors Eq.~(\ref{Localtens4}) in two steps. Consider first fixing the index $\textbf{j}$ to the same value $\textbf{j}^*$ on each link. Then, summing over the indices $a,b$ on each link produces the term $c_{\textbf{j}^*}[X(\textbf{j}^*)+\mathrm{trans}.]$. Note that this sum is greatly simplified because, in order to get a non-zero contraction, fixing the indices $a,b$ on one link also fixes them on all other links. Now we sum over $\textbf{j}^*$ to get $T_\Phi$ as defined in Eq.~(\ref{tphi}), up to a constant factor $n^2N$. We emphasize that the same tensor $B_\Phi$ is placed on each lattice site, so the tensor network is translationally invariant.

(ii) The tensor $A_\Phi$ representing the state $|\Phi\rangle$ is constructed as $A_\Phi=B_\Phi C$. Graphically,
$$
\parbox{3cm}{\includegraphics[width=2.8cm]{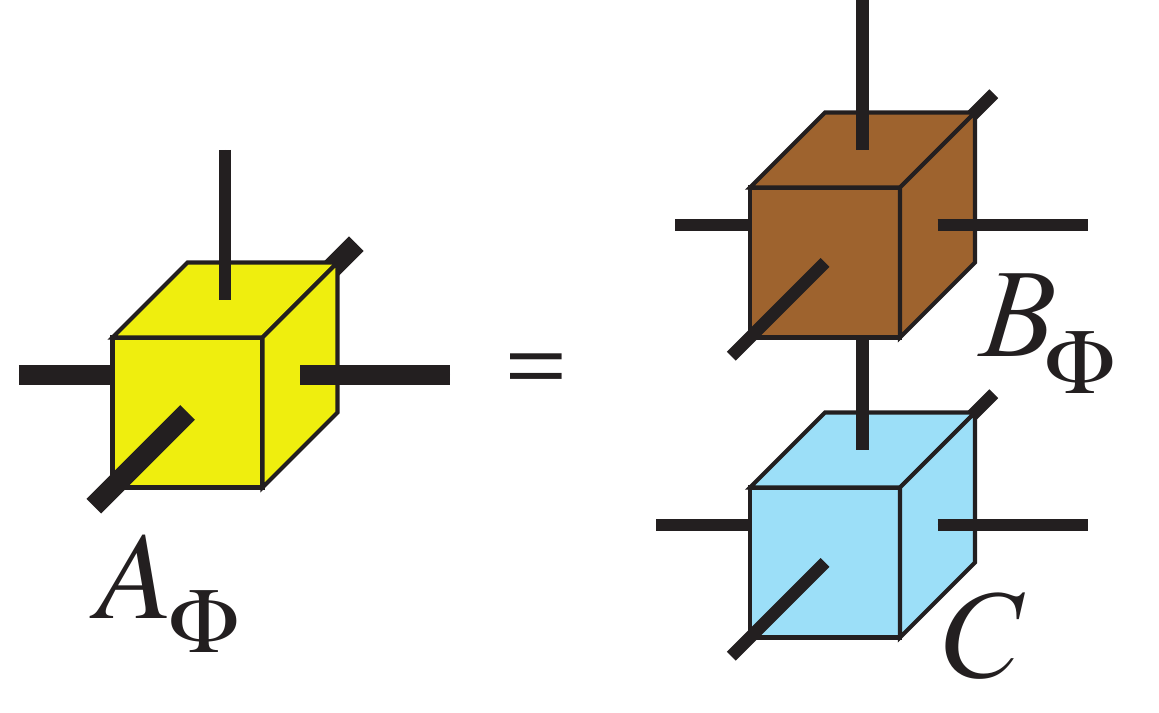}}.
$$
\color{black}
With the form Eq.~(\ref{Localtens4}) of the tensors $B_\Phi$, those tensors commute with the action of a local Pauli $X$-operator on the physical legs,
$$
\parbox{3.3cm}{\includegraphics[width=3.3cm]{LocTensB}}.
$$
With this relation, and since the symmetries Eq.~(2) [m] hold for the cluster state tensor $C$, we have,
$$
\parbox{8.1cm}{\includegraphics[width=8cm]{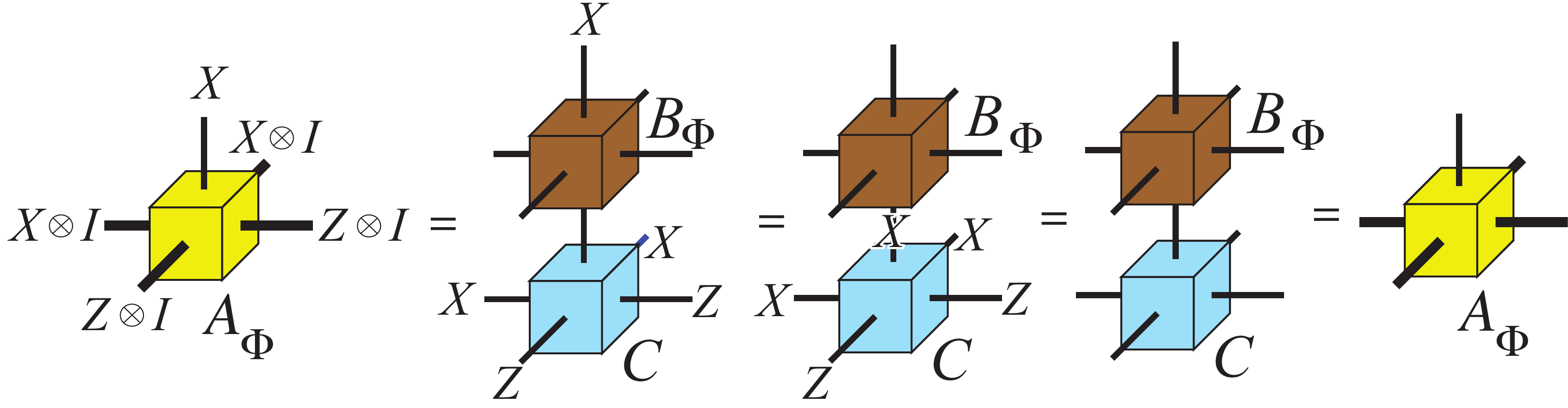}}.
$$
Thus, the tensors $A_\Phi$ describing the state $|\Phi\rangle$ share the first of the symmetries in Eq.~(2) [m] with the cluster state tensors $C$. The same holds for the remaining symmetries of Eq.~(2) [m], and the proof is analogous.

\section{Correlations in virtual space}

\begin{figure}[htb]
	\begin{center}
		\includegraphics[width=6.5cm]{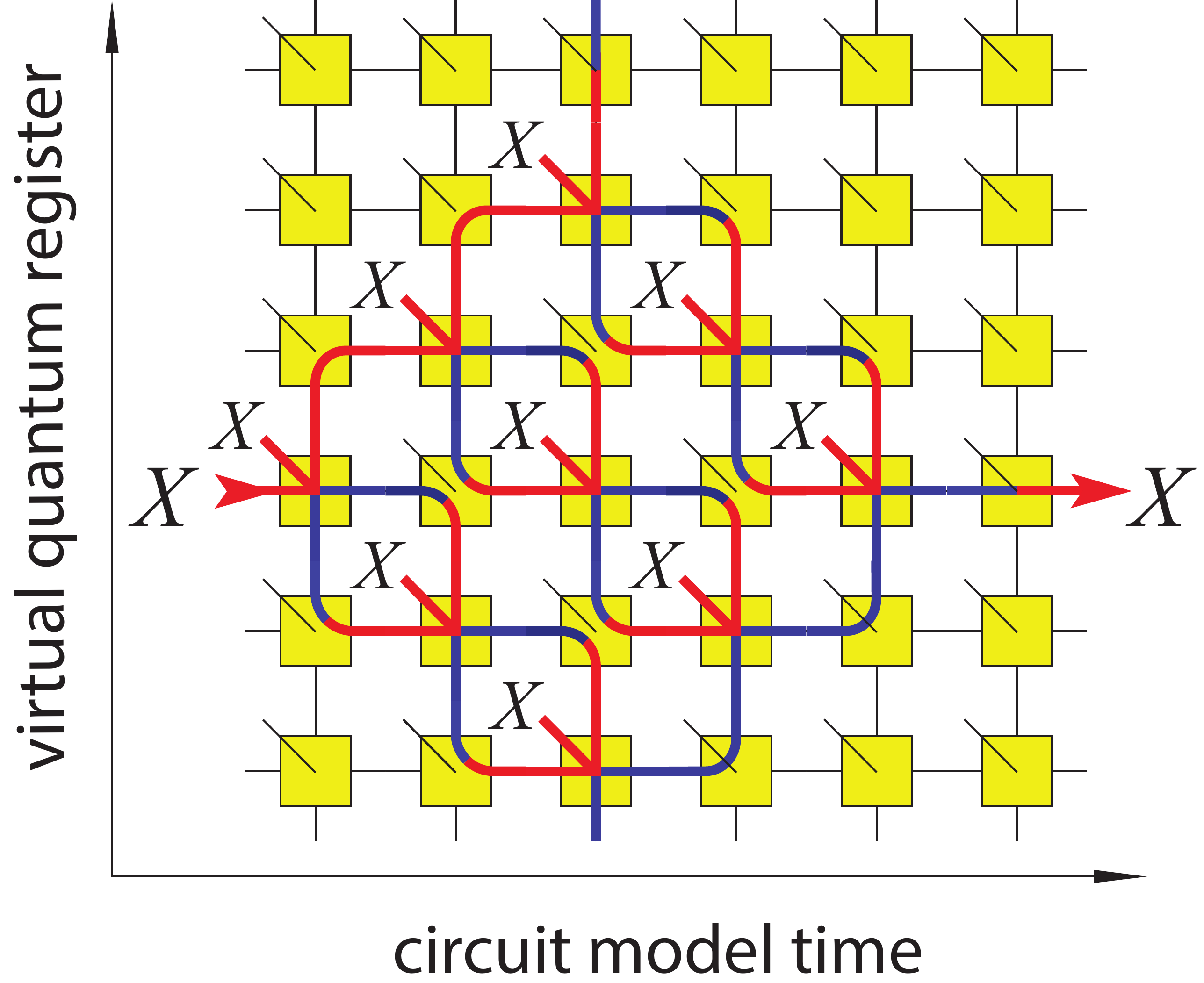}
		\caption{\label{QCAf2} Symmetry-protected quantum correlations complementary to the symmetries in Fig.~1. The red (blue) lines represent the action of Pauli operators $X$ ($Z$) on the virtual legs of PEPS tensors, such that matching pairs cancel. Symmetries displayed in both figures and their vertical translates generate a symmetry group $\tilde{G}=\mathbb{Z}_2^n\times\mathbb{Z}_2^n$. The boundary conditions in the vertical direction are periodic.}
	\end{center}
\end{figure}

As an additional guide to the proof of Lemma~2 in the main text,  a graphical representation of an $X$-type correlation in virtual space is provided in Fig.~\ref{QCAf2}.

\section{Canonical form for the quasi-1D block tensors ${\cal{A}}_\Phi$}

We briefly expand on the importance of the canonical form for computation. Given a quantum state, there are infinitely many ways to represent it as an MPS. Using this freedom, one can always choose an MPS tensor such that it is in the canonical form \cite{MPS}. The canonical form of an MPS is a powerful tool that allows one to relate mathematical properties of the MPS tensor to physical properties of the state. For example, an MPS has a finite correlation length if and only if a certain linear map (to be introduced in the next section) has a unique fixed point. We require this uniqueness for the computational primitive of oblivious wire, upon which all non-trivial gates are based; see \cite{SPTO1} and  the next section. It is guaranteed it in the canonical form since every state in the cluster phase has a finite correlation length. 

In the main text and Section~\ref{GenA}, we construct a PEPS representation $A_\Phi$ of a state ${|\Phi\rangle}$ in the cluster phase by stacking a tensor network representation of the circuit $U_\Phi$ on top of the cluster state PEPS. While this is legitimate PEPS representation, it may not lead to an MPS in canonical form upon blocking into the tensor $\mathcal{A}_\Phi$. Hence, we need to go through the extra step of showing that our results also hold in canonical form in order to use its properties when constructing a computational scheme.

\subsection{Proof of Lemma~2 in canonical form} \label{app:lemma3}

\noindent

Here we use the quasi-1D picture to show that Lemma~2 [m] holds even when our block MPS tensor $\mathcal{A}_\Phi$ is put into canonical form.\smallskip

\emph{Proof of Lemma 2 in canonical form.} 
The proof proceeds by considering the quasi-1D SPT phase containing the 2D cluster state protected by the group $\tilde{G}$ of cone-like symmetries generated by the symmetries shown in Fig.~(1) [m] and Fig.~(\ref{QCAf2}), and their translates. For every state in this phase, Eqs.~(7) [m] and~(8) [m] are guaranteed to hold for the canonical MPS representation of that state, as stated in Theorem 1 of Ref.\cite{Bartl}. To complete the proof, we need only show that every state in the cluster phase is also in this quasi-1D phase.

Consider a state ${|\Phi\rangle}=U_\Phi {|\mathcal{C}\rangle}$ in the cluster phase. The circuit $U_\Phi$ is defined to be symmetric under the line symmetries defining the cluster phase. It turns out that this guarantees that it is also symmetric under the cone symmetries defining the group $\tilde{G}$, which is a strictly larger group than the group $G$ formed by the line symmetries. This is because $U_\Phi$ is a local circuit of constant depth with a 2D notion of locality. 

In the proof of Lemma~1 [m] we identified three types of Pauli operators that commute with the stripe symmetries. While the local operators shown in Fig.~2 (b) [m] commute with all elements of $\tilde{G}$, non-local operators like that in Fig.~2 (c) [m] do not. But, because of the local structure of $U_\Phi$, the non-local operators will not appear in its expansion (Eq.~(3) [m]). So every $U_\Phi$ which commutes with the stripe symmetries will also commute with all elements of $\tilde{G}$. The circuit $U_\Phi$ is local in the 2D sense, so it also satisfies the weaker notion of 1D block locality. Hence, ${|\Phi\rangle}$ is connected to the cluster state via a local circuit of constant depth which commutes with $\tilde{G}$, and it is thus in the quasi-1D SPT phase protected by $\tilde{G}$. $\Box$

\subsection{Operators $\mathcal{C}(\mathfrak{i})$ in canonical form}

Now that we have shown that Eqs.~(7) [m] and~(8) [m] hold in the canonical form, we would like to identify the precise form of the operators $\mathcal{C}(\mathfrak{i})$ therein. In particular, we must confirm that the relations given in Eq. (11) [m] still hold in canonical form, as we didn't establish this of Eq.~(9) [m] which was used to derive  Eq. (11) [m]. We again use Theorem 1 of Ref.~\cite{Bartl}, which gives an explicit recipe for determining $\mathcal{C}(\mathfrak{i})$. 

Denote by $u(g)$ the representation of $\tilde{G}$ given by the cone-like symmetry operators in Figs.~1 [m] and \ref{QCAf2}.
Depending on the binary vector $\mathfrak{i}$, the action of $u(g)$ on $\mathcal{A}_\Phi(\mathfrak{i})$ yields a sign $\chi_\mathfrak{i}(g)=\pm 1$ according to $u(g)\ket{\mathfrak{i}}=\chi_\mathfrak{i}(g)\ket{\mathfrak{i}}$. Then we have~\cite{Bartl}
\begin{equation}\label{BlockEq}
\mathcal{C}(\mathfrak{i})V(g)=\chi_\mathfrak{i}(g)\: V(g)\mathcal{C}(\mathfrak{i}),\:\:\forall g\in\tilde{G}.
\end{equation} 
Therein, $V(g)$ are the Pauli operators acting in the virtual space associated to $u(g)$ as in Figs.~1 [m] and \ref{QCAf2}. Writing $\mathcal{C}(\mathfrak{i}) \sim \prod_{k \in K} \mathcal{C}[k]^{i_k}$ as in the main text, Eq.~(\ref{BlockEq}) can be used to determine $\mathcal{C}[k]$. Then, one can straightforwardly  verify that the operators given in Eq.~(11) [m] satisfy Eq.~(\ref{BlockEq}), and hence they appear in the canonical form of the tensor $\mathcal{A}_\Phi$, as desired.\smallskip

We restate Eq.~(11)~[m] here for later use in Section~\ref{TmProof}, now valid also in canonical form,
\begin{subequations}
\label{QCAsum2}
\begin{align}
\label{QCAZ}
\mathcal{C}[(1,l)]&=Z_l,\\
\label{QCAX}
\mathcal{C}[(n,l)]&=X_l,\\
\label{QCAZXZ}
\mathcal{C}[(2,l)]&=Z_{l-1}X_lZ_{l+1}.
\end{align}
\end{subequations}

\section{Computation in SPT phases}

\subsection{Review of techniques for quantum computation in SPT phases}\label{Rev}

Here we review the basic techniques for computing in 1D SPT phases from \cite{SPTO1}. The proof of Theorem~1, which is provided in the next section, substantially relies on these techniques. This section is only intended as an intuitive guide to the main techniques employed in \cite{SPTO1}. It does not replace \cite{SPTO1} as technical background. 

In the 1D scenario discussed here, we deal with a chain of $N$-level spins, without any substructure. We discuss two techniques for MBQC in SPT phases, namely ``oblivious wire'', and ``logical gates by averaging over measurement outcomes''.

{\em{Oblivious wire.}} The purpose of oblivious wire is to decouple the logical subsystem of the virtual space from the junk subsystem. An elementary segment ${\cal{L}}$ of oblivious wire is implemented by (i) measuring a number $m$ of successive spins in the symmetry protected basis, (ii) propagating the logical byproduct operator $\mathcal{C}(\mathfrak{i})$ forward to the end of the computation, and (iii) forgetting the measurement outcomes.

The resulting action on the virtual space is $I\otimes {\cal{L}}$, with
$$
{\cal{L}}(\rho) = \sum_{i=0}^{N-1} \mathcal{B}_\Phi(i) \rho \mathcal{B}_\Phi(i)^\dagger.
$$
Note that the tensors ${\cal{B}}_\Phi$ are location-independent, cf. Section~\ref{GenA}. Here we finally use the property of translation-invariance of the physical setup.

Since the ground state anywhere in the SPT phase has finite correlation length, the MPS tensors ${\cal{B}}_\Phi(i)$ are injective \cite{Bartl}, \cite{Bartl2}, and ${\cal{L}}$ therefore has a unique fixed point. Note that these implications hold only for tensors in the canonical form, hence why we require the proof in Appendix B. Denote by $\lambda_1$ the second-largest eigenvalue of ${\cal{L}}$, and $\xi:=-1/\ln \lambda_1$ the corresponding correlation length. Then, for $m \gg \xi$,
$$
{\cal{L}}^{m}(\rho) \rightarrow \nu_\rho \rho_\text{fix},
$$
where $\rho_\text{fix}$ is the unique fixed point of ${\cal{L}}$ and $\nu_\rho \in \mathbb{R}_+$. Oblivious wire applied to a state $\tau$ defined on the entire virtual space decouples the logical from the junk part,
$$
{\cal{L}}^m(\tau) \rightarrow \sigma \otimes \rho_{\text{fix}}.
$$

{\em{Logical gates.}} Small rotations $\sim d\alpha$ are implemented by measuring slightly off the symmetry protected basis, followed by oblivious wire. The reason for implementing only small rotation angles between two pieces of oblivious wire is that unitarity on the logical subspace is violated at second order in $d\alpha$ \cite{SPTO1}. 

To realize a logical unitary, we measure in the basis
\begin{equation}\label{Mbas}
\begin{array}{rcl}
|i\rangle' &=& |i\rangle +  e^{i\delta}d\alpha\, |j\rangle,\\
|j\rangle' &=& |j\rangle -  e^{i\delta}d\alpha\, |i\rangle,\\
\end{array}
\end{equation}
and $|r\rangle' = |r\rangle$ for $r\neq i,j$. Therein, $\{|i\rangle,\, i=0,..,N-1\}$ is the symmetry-respecting basis. Then, the logical operation
\begin{equation}\label{Rot}
U(\delta,d\alpha) =\exp\left(i d\alpha \frac{e^{-i\delta}\nu_{ji}\mathcal{C} -e^{i\delta}\nu_{ji}^*\mathcal{C}^\dagger}{i} \right)
\end{equation}
is implemented, for small angles $d\alpha$. Therein,
\begin{equation}\label{CijDef}
\mathcal{C}=\mathcal{C}(i)^{-1}\mathcal{C}(j),
\end{equation} for $0 \leq i,j \neq N-1$ and $i\neq j$,
and the constant $\nu_{ij} \in \mathbb{C}$ is given by
$$
\lim_{m\rightarrow \infty} {\cal{L}}^m( \mathcal{B}_{\Phi}(i) \rho_\text{fix} \mathcal{B}_{\Phi}(j)^\dagger) = \nu_{ij} \, \rho_\text{fix}.
$$
$U(\delta,d\alpha)$ is implemented after accounting for the byproduct operator through forward propagation, and subsequent averaging over the measurement outcomes. Finite rotation angles are accumulated by repetition.

The constants $\nu_{ij}\in \mathbb{C}$ affect the gate operation and vary across the phase. They need to be measured prior to computation. By translation invariance they are the same for all sites. \smallskip

Finally, a weak measurement of the observable $\mathcal{C}$ is performed by measuring in the basis
$$
\left\{\frac{|i\rangle + |j\rangle}{\sqrt{2}}, \frac{|i\rangle - |j\rangle}{\sqrt{2}}, |r\rangle,\;  \forall r\neq i,j \right\}
$$
A near-projective measurement of $\mathcal{C}$ is achieved by repeating the above procedure many times. For details and the changeover between logical unitary and logical measurement see \cite{SPTO1}. 

\subsection{Proof of Theorem~1}
\label{TmProof}

The proof of Theorem~1 is similar to that of Theorem~2 of \cite{SPTO1}, see Sections III - VI therein. There is one additional complication, however. In the present setting there are two distinct notions of locality, namely ``block locality'' on which the effective 1D SPT order hinges, and ``single-site locality'' on which MBQC hinges. The latter is more stringent: only single site measurements are available to MBQC, not block-local ones. It needs to be shown that quantum computational universality persists under this restriction of measurement bases.  

For concreteness, in the 1D case we could choose to mix any pair of basis states $|i\rangle$ and $|j\rangle$ in Eq.~(\ref{Mbas}), to construct a computationally useful measurement basis. In the present 2D case, the corresponding linear combinations of block states $|\mathfrak{i}\rangle$ and $|\mathfrak{j}\rangle$  are constrained by the requirement of single-site locality.\medskip 

{\em{Proof of Theorem~1.}} The proof proceeds in 3 steps, which address unitary gates, universality, and measurement and initialization. 

(i) {\em{Unitary gates.}} Consider all but one qubit $k$ in a given block being measured in the symmetry protected $X$-basis, while qubit $k$ is measured in the basis ${\cal{B}}(\delta,d\alpha)$ spanned by
\begin{equation}\label{TiltBas}
|+\rangle' = |+\rangle + e^{i\delta} d\alpha |-\rangle,\; |-\rangle' = |-\rangle - e^{i\delta}d\alpha |+\rangle.
\end{equation}
These measurements are followed by oblivious wire. 

\begin{figure}[htb]
	\begin{center}
		\begin{tabular}{ll}
			(a) & (b) \\
			\includegraphics[width=4cm]{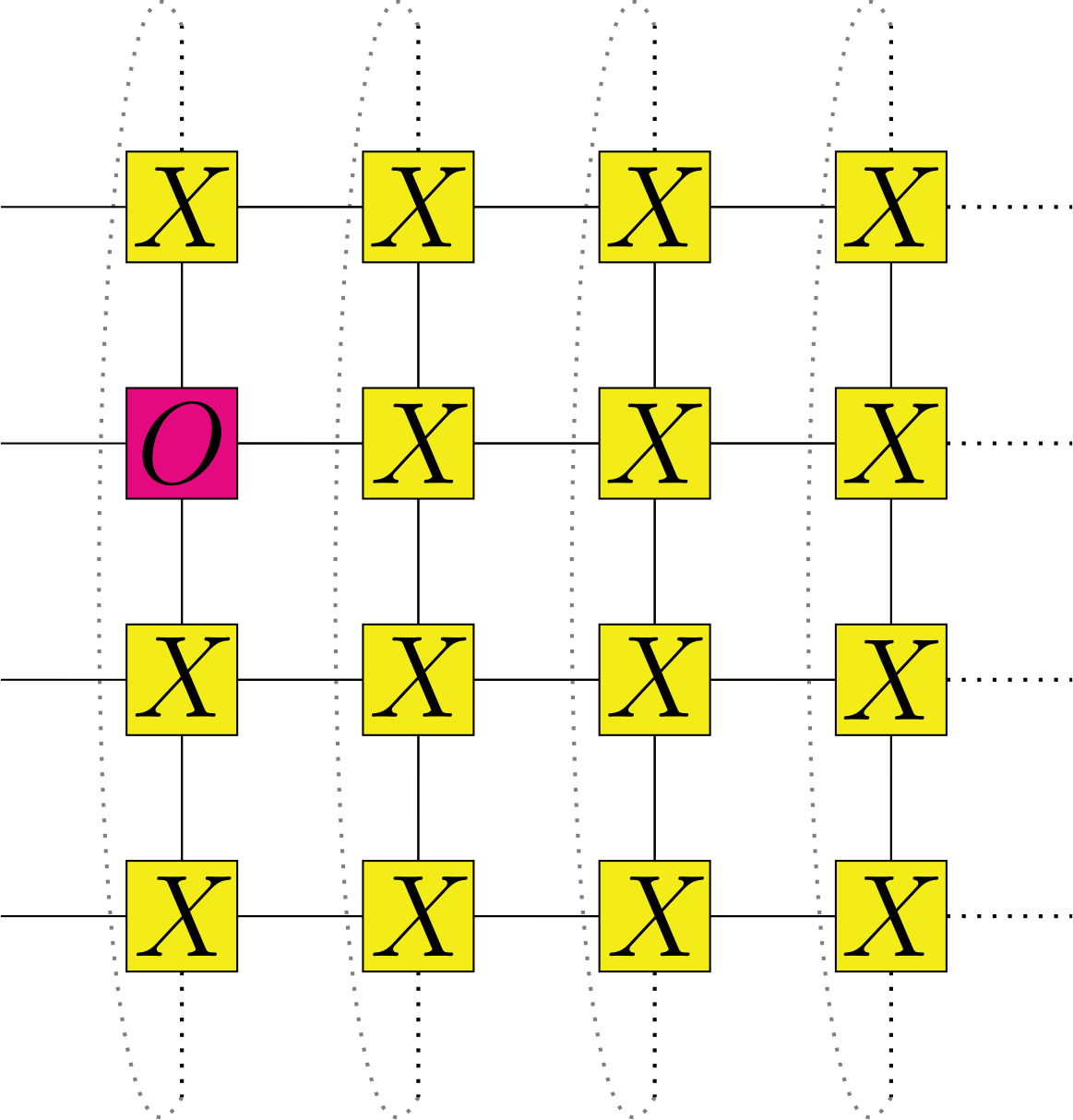} & \includegraphics[width=4cm]{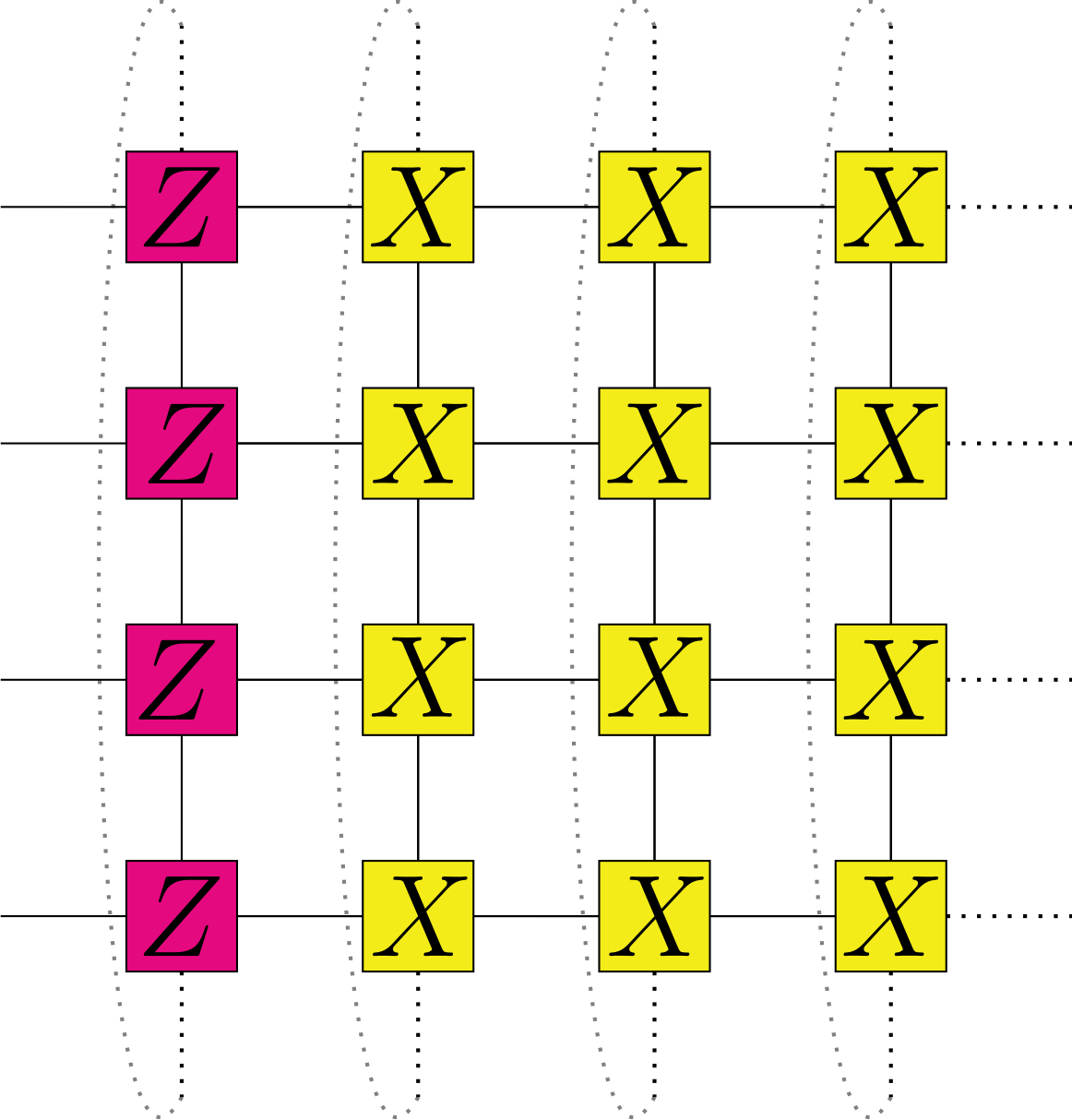}
		\end{tabular}
		\caption{\label{Mpat} Illustration for the proof of Theorem~1. (a) Measurement patterns for unitary gates and measurements. (b) Simplified pattern for initialization by measurement.}
	\end{center}
\end{figure}

In the following, we establish that (1) single site measurements in the basis ${\cal{B}}(\delta,d\alpha)$ are sufficient to implement logical quantum gates, and (2) the resulting gate set is universal.

(1) We consider the conditional logical transformation implemented given certain measurement outcomes. Denote $\overline{\mathfrak{i}}^{(k)}:= \mathfrak{i} +\mathfrak{e}_k\mod 2$,  for any given site $k$ under consideration. Therein, $\mathfrak{e}_k$ is the $n^2$-component binary vector with an entry 1 in position $k$, and 0 everywhere else. $T_{\mathfrak{i} \cup \overline{\mathfrak{i}}^{(k)}}$ is the conditional logical transformation applied if either the outcome $\mathfrak{i}$ or $\overline{\mathfrak{i}}^{(k)}$ was obtained, and $p\big(\mathfrak{i} \cup \overline{\mathfrak{i}}^{(k)}\big)$ is the probability for obtaining the outcome $\mathfrak{i}$ or $\overline{\mathfrak{i}}^{(k)}$. In analogy with Eq.~(\ref{CijDef}) for the 1D case, we define the operators 
\begin{equation}
\mathcal{C}_{\mathfrak{ij}}:=C(\mathfrak{i})^{-1}C(\mathfrak{j}).
\end{equation}
We are interested in particular in the operators $\mathcal{C}_{\mathfrak{i}\overline{\mathfrak{i}}^{(k)}}$ which result from measuring qubit $k$ of the given block in a symmetry-breaking local basis of Eq.~(\ref{TiltBas}).  From Eq.~(10) [m] it follows that 
$$
\mathcal{C}_{\mathfrak{i}\overline{\mathfrak{i}}^{(k)}} = \mathcal{C}[k],\;\; \forall \mathfrak{i}.
$$
We thus find the following form for the logical operation $T_{\mathfrak{i} \cup \overline{\mathfrak{i}}^{(k)}}$ resulting from the measurement in the basis Eq.~(\ref{TiltBas}), up to linear order in $d\alpha$,
\begin{equation}\label{T_i}
\begin{array}{rcl}
 p\left(\mathfrak{i} \cup \overline{\mathfrak{i}}^{(k)}\right)  T_{\mathfrak{i} \cup \overline{\mathfrak{i}}^{(k)}} &=& \displaystyle{(\nu_{\mathfrak{ii}}+\nu_{\overline{\mathfrak{i}}^{(k)} \overline{\mathfrak{i}}^{(k)}})I +}\vspace{1mm}\\
&& \displaystyle{+ i d\alpha \frac{e^{-i\delta}\nu_{\mathfrak{i}\overline{\mathfrak{i}}^{(k)}}\mathcal{C}[k] -e^{i\delta}\nu_{\mathfrak{i} \overline{\mathfrak{i}}^{(k)}}^*\mathcal{C}[k]^\dagger}{i} .}
\end{array}
\end{equation}
Therein, the overall byproduct operator $\mathcal{C}(\mathfrak{i})$ has been separated out.  The constants $\nu_{\mathfrak{i}\overline{\mathfrak{i}}^{(k)}}$ are given by the relation
$$
\lim_{m\rightarrow \infty} {\cal{L}}^m( \mathcal{B}_{\Phi}(\mathfrak{i}) \rho_\text{fix} \mathcal{B}_{\Phi}(\mathfrak{j})^\dagger) = \nu_{\mathfrak{ij}} \, \rho_\text{fix},
$$
with $\rho_{\text{fix}}$ the unique fixed point state of the junk subsystem. 
The uniqueness of $\rho_\text{fix}$ is guaranteed by Lemma~2 in canonical form, as proved in Section~\ref{app:lemma3}.

As in \cite{SPTO1}, logical gates are implemented by ``forgetting'' the measurement outcomes $\mathfrak{i}$ after accounting for the corresponding byproduct operator. That is, we implement the weighted probabilistic average of the gates $T_{\mathfrak{i}\cup \overline{\mathfrak{i}}^{(k)}}$ described in Eq.~(\ref{T_i}). These gates are $U(\delta,d\alpha)=\sum_{\mathfrak{i}|\,\mathfrak{i}_k=0}p(\mathfrak{i} \cup \overline{\mathfrak{i}}^{(k)})  T_{\mathfrak{i} \cup \overline{\mathfrak{i}}^{(k)}}$, where $\mathfrak{i}_k$ is the $k$-th component of $\mathfrak{i}$. With $\sum_{\mathfrak{i}}\nu_\mathfrak{ii} =1$ \cite{SPTO1}, we obtain
\begin{equation}\label{Ugate}
U(\delta,d\alpha) = I + i d\alpha \frac{e^{-i\delta}\nu[k]\mathcal{C}[k] -e^{i\delta}\nu[k]^*\mathcal{C}[k]^\dagger}{i},
\end{equation}
with
$\nu[k]:=\sum_{\mathfrak{i}|\,\mathfrak{i}_k=0}\nu_{\mathfrak{i}\overline{\mathfrak{i}}^{(k)}}$. These gates are indeed unitary up to linear order in $d\alpha$. The angle $d\alpha$ therefore has to be chosen small. Finite rotation angles are accumulated by repetition.

At special points in the phase, reachable by fine-tuning, the conversion rate of measurement angle to rotation angle is exactly zero. For those ground states we cannot establish universal computational power. Hence the restriction to ``all ground states except a possible set of measure zero'' in Theorem~1.\medskip

(2) Now we make special choices for the site $k$. First, for $k=(1,l)$, with Eq.~(\ref{QCAZ}), for a suitable angle $\delta$ the gate action is
$
U_{(1,l)}(d\alpha) = \exp\left(2id\alpha\, |\nu[(1,l)]| Z_l \right),
$
i.e., an infinitesimal rotation about the $Z_l$-axis is performed.

Second, choosing $k=(n,l)$, leads to gates $U_{(n,l)}=\exp(2i|\nu[(n,l)]| d \alpha X_l)$ due to Eq.~(\ref{QCAX}). Finally, choosing $k=(2,l)$ leads to gates
\begin{equation}\label{UK}
U_{(2,l)} = \exp(2i |\nu[(2,l)]|d \alpha Z_{l-1} X_l Z_{l+1}),
\end{equation}
in accordance with Eqs.~(\ref{Ugate}) and~(\ref{QCAZXZ}).
Again, finite rotation angles are accumulated by repetition.
	
(ii) {\em{Universality.}} We show that the local $X$ and $Z$-rotations, and the entangling gates of Eq.~(\ref{UK}) form a universal set for $n/2$ qubits. After initialization, all logical qubits are in a $Z$-eigenstates. Through the available local rotations, they can be rotated into the logical state $|+\rangle$, up to the action of byproduct operators in ${\cal{P}}_n$. Thereafter, the state of the odd-numbered qubits remains unchanged throughout. Only the local rotations for even-numbered qubits, and the rotations Eq.~(\ref{UK}) for odd $k$ are subsequently used. Because  the odd-numbered  qubits are frozen to $|+\rangle$, the latter gates become
$$
U_{(2,l)} \cong \exp(i \beta Z_{l-1} Z_{l+1}),\;\; \text{for} \; l\; \text{even}.
$$ 
Together these gates are universal for the even-numbered qubits. 

(iii) {\em{Measurement and initialization.}} With Eq.~(\ref{QCAZ}), for weak measurements of  a logical qubit $l$ in the $Z$-basis, the measurement pattern of Fig.~\ref{Mpat}a, with $O=Z$, is performed. For a near-projective measurement of the logical subsystem of the virtual space, this pattern is repeated on a large number of consecutive blocks of physical qubits. 

Initialization can be performed by measurement. To simplify the procedure, all logical qubits can be measured in the $Z$-basis simultaneously by the measurement pattern in Fig.~\ref{Mpat}b. 
$\Box$


\begin{thebibliography}{99}

\bibitem{Doh}
A.C. Doherty, S.D. Bartlett, Phys. Rev. Lett. \textbf{103}, 020506 (2009).

\bibitem{M1}
A. Miyake, Phys. Rev. Lett. \textbf{105}, 040501 (2010).

\bibitem{Bartl}
D.V. Else, I. Schwarz, S.D. Bartlett and A.C. Doherty, Phys. Rev. Lett. \textbf{108}, 240505 (2012).

\bibitem{MM2}
J. Miller and A. Miyake, Phys. Rev. Lett. 114, 120506 (2015).

\bibitem{SPTO2}
D.T. Stephen, D.-S. Wang, A. Prakash, T.-C. Wei, R. Raussendorf, Phys. Rev. Lett. \textbf{119}, 010504  (2017).

\bibitem{SPTO1}
R. Raussendorf, D.-S. Wang, A. Prakash, T.-C. Wei, D.T. Stephen, Phys. Rev. A \textbf{96}, 012302,  (2017).

\bibitem{Darmawan}
A.S. Darmawan, G.K. Brennen, S.D. Bartlett, New J. Phys. \textbf{14}, 013023 (2012).

\bibitem{HuaWei}
C.-Y. Huang, M.A. Wagner, and T.-C. Wei, arXiv:1605.08417.

\bibitem{RB01}
R. Raussendorf and H.-J. Briegel, Phys. Rev. Lett. \textbf{86}, 5188 (2001).

\bibitem{GW}
Z.C. Gu and X.G. Wen, Phys. Rev. B \textbf{80}, 155131 (2009).

\bibitem{Wen1}
X. Chen, Z.C. Gu, and X.G. Wen, Phys. Rev. B \textbf{82}, 155138 (2010).

\bibitem{Wen2}
X. Chen, Z.C. Gu, Z.X. Liu, X.G. Wen, Phys. Rev. B \textbf{87}, 155114 (2013).

\bibitem{Zitz}
H. Niggemann, A. Kl{\"u}mper, and J. Zittartz, Z. Phys. B \textbf{104}, 103 (1997).

\bibitem{WeiHua}
T.-C. Wei and C.-Y. Huang, Phys. Rev. A \textbf{96}, 032317 (2017).

\bibitem{Poul}
H. Poulsen Nautrup and T.C. Wei, Phys. Rev. A \textbf{92}, 052309 (2015).

\bibitem{MM1}
J. Miller and A. Miyake, NPJ Quantum Information \textbf{2}, 16036 (2016).

\bibitem{CPT}
Y. Chen, A. Prakash, T.C. Wei, 	Phys. Rev. A \textbf{97}, 022305 (2018).

\bibitem{MM3}
J. Miller and A. Miyake, arXiv:1612.08135.


\bibitem{HuangChen}
Y. Huang, X. Chen, Phys. Rev. B \textbf{91}, 195143 (2015).

\bibitem{Yosh}
B. Yoshida, Phys. Rev. B \textbf{93}, 155131 (2016).

\bibitem{Bartl3}
S. Roberts, B. Yoshida, A. Kubica, S.D. Bartlett, Phys. Rev. A \textbf{96}, 022306 (2017).

\bibitem{Bartl2}
D.V. Else, S.D. Bartlett, A.C. Doherty, New J. Phys. \textbf{14}, 113016 (2012).

 
\bibitem{MPS}
D. Perez-Garcia, F. Verstraete, M.M. Wolf, J.I. Cirac, Quant. Inf. Comput. \textbf{7}, 401 (2007).

\bibitem{Fitz}
A. Mantri, T.F. Demarie, J.F. Fitzsimons, arXiv:\-1607.\-00758.

\bibitem{Eis}
D. Gross and J. Eisert, Phys. Rev. Lett. \textbf{98}, 220503 (2007).

\bibitem{KT}
A. Kapustin, R. Thorngren, In: Algebra, Geometry, and Physics in the 21st Century. Progress in Mathematics \textbf{324}. Birkh{\"a}user Verlag, Cham, Switzerland, 177-202.

\bibitem{TIQC}
R. Raussendorf, Phys. Rev. \textbf{72}, 052301 (2005).

\bibitem{W1}
D.M. Schlingemann,  H. Vogts and R.F. Werner, J. Math. Phys. \textbf{49}, 112104 (2008).

\bibitem{W2}
D. Gross, V. Nesme, H. Vogts, R.F. Werner, Commun. Math. Phys. \textbf{310}, 419 (2012).

\bibitem{W3}
C. Cedzich, T. Geib, F.A. Gr{\"u}nbaum, C. Stahl, L. Velazquez, A.H. Werner, R.F. Werner, Ann. Henri Poincar{\'e} \textbf{19},  325 (2018). 

\bibitem{W4}
J.I. Cirac, D. Perez-Garcia, N. Schuch, F. Verstraete, J. Stat. Mech. 083105 (2017).

\end{thebibliography}
\end{document}